\documentstyle[preprint,aps,psfig,floats,tighten]{revtex}

\newcommand{\dzero}     {D\O}
\newcommand{\met}       {\mbox{$\not\!\!E_T$}}
\newcommand{\mdeta}     {\mbox{$|\eta^{\rm det}|$}}
\newcommand{\deta}      {\mbox{$\eta^{\rm det}$}}
\newcommand{\dtheta}    {\mbox{$\theta^{\rm det}$}}
\newcommand{\rar}       {\rightarrow}
\newcommand{\rargap}    {\mbox{ $\rightarrow$ }}
\newcommand{\ejm}       {\mbox{$e$+jets/$\mu$}}
\newcommand{\mjm}       {\mbox{$\mu$+jets/$\mu$}}
\newcommand{\ttbar}     {\mbox{$t\bar{t}$}}
\newcommand{\bbbar}     {\mbox{$b\bar{b}$}}
\newcommand{\ppbar}     {\mbox{$p\bar{p}$}}
\newcommand{\Vtb}       {\mbox{$V_{tb}$}}
\newcommand{\Vtbsq}     {\mbox{$\left|V_{tb}\right|^2$}}
\newcommand{\comphep}   {\sc{c}\rm{omp}\sc{hep}}
\newcommand{\herwig}    {\sc{herwig}}
\newcommand{\pythia}    {\sc{pythia}}
\newcommand{\jetset}    {\sc{jetset}}
\newcommand{\geant}     {\sc{geant}}

\begin{document}

\preprint{Fermilab-Pub-00/188-E}

\title{\hfill \break \hfill \break
       Search for Electroweak Production of Single Top Quarks \\
       in {\ppbar} Collisions \vspace{0.2 in}}

\author{\centerline{The {\dzero} Collaboration
        \thanks{PACS numbers: 14.65.Ha, 12.15.Ji, 13.85.Qk}}}

\address{\centerline{Fermi National Accelerator Laboratory,
                     Batavia, Illinois 60510} \vspace{0.5 in}}

\maketitle

\begin{abstract}
We present a search for electroweak production of single top quarks in
the electron+jets and muon+jets decay channels. The measurements use
$\approx90$~pb$^{-1}$ of data from Run~1 of the Fermilab Tevatron
collider, collected at 1.8~TeV with the {\dzero} detector between 1992
and 1995. We use events that include a tagging muon, implying the
presence of a $b$ jet, to set an upper limit at the $95\%$ confidence
level on the cross section for the $s$-channel process
${\ppbar}{\rargap}tb+X$ of 39~pb. The upper limit for the $t$-channel
process ${\ppbar}{\rargap}tqb+X$ is 58~pb.
\end{abstract}

%\pacs{PACS numbers: 14.65.Ha, 12.15.Ji, 13.85.Qk}

\clearpage

% LIST_OF_AUTHORS.TEX              8/10/00

\centerline{{\bf The {\dzero} Collaboration}}
\vspace{0.1 in}
\small

\begin{center}
B.~Abbott,$^{50}$ M.~Abolins,$^{47}$ V.~Abramov,$^{23}$
B.S.~Acharya,$^{15}$ D.L.~Adams,$^{57}$ M.~Adams,$^{34}$
G.A.~Alves,$^{2}$ N.~Amos,$^{46}$ E.W.~Anderson,$^{39}$
M.M.~Baarmand,$^{52}$
% ...
\end{center}

\vspace{0.1 in}

\centerline{$^{1}$Universidad de Buenos Aires, Buenos Aires,
                  Argentina}       
\centerline{$^{2}$LAFEX, Centro Brasileiro de Pesquisas F{\'\i}sicas,
                  Rio de Janeiro, Brazil}
\centerline{$^{3}$Universidade do Estado do Rio de Janeiro,
                  Rio de Janeiro, Brazil}
\centerline{$^{4}$Institute of High Energy Physics, Beijing,
                  People's Republic of China}
\centerline{$^{5}$Universidad de los Andes, Bogot\'{a}, Colombia}
\centerline{$^{6}$Charles University, Prague, Czech Republic}
\centerline{$^{7}$Institute of Physics, Academy of Sciences, Prague,
                  Czech Republic}
\centerline{$^{8}$Universidad San Francisco de Quito, Quito, Ecuador}
\centerline{$^{9}$Institut des Sciences Nucl\'eaires, IN2P3-CNRS,
                  Universite de Grenoble 1, Grenoble, France}
\centerline{$^{10}$CPPM, IN2P3-CNRS, Universit\'e de la
                   M\'editerran\'ee, Marseille, France}
\centerline{$^{11}$LPNHE, Universit\'es Paris VI and VII, IN2P3-CNRS,
                   Paris, France}
\centerline{$^{12}$DAPNIA/Service de Physique des Particules, CEA,
                   Saclay, France}
\centerline{$^{13}$Panjab University, Chandigarh, India}
\centerline{$^{14}$Delhi University, Delhi, India}
\centerline{$^{15}$Tata Institute of Fundamental Research, Mumbai,
                   India}
\centerline{$^{16}$Seoul National University, Seoul, Korea}
\centerline{$^{17}$CINVESTAV, Mexico City, Mexico}
\centerline{$^{18}$FOM-Institute NIKHEF and University of
                   Amsterdam/NIKHEF, Amsterdam, The Netherlands}
\centerline{$^{19}$University of Nijmegen/NIKHEF, Nijmegen, The
                   Netherlands}
\centerline{$^{20}$Institute of Nuclear Physics, Krak\'ow, Poland}
\centerline{$^{21}$Institute for Theoretical and Experimental
                   Physics, Moscow, Russia}
\centerline{$^{22}$Moscow State University, Moscow, Russia}
\centerline{$^{23}$Institute for High Energy Physics, Protvino,
                   Russia}
\centerline{$^{24}$Lancaster University, Lancaster, United Kingdom}
\centerline{$^{25}$University of Arizona, Tucson, Arizona 85721}
\centerline{$^{26}$Lawrence Berkeley National Laboratory and
                   University of California, Berkeley, California
                   94720}
\centerline{$^{27}$University of California, Davis, California 95616}
\centerline{$^{28}$California State University, Fresno, California
                   93740}
\centerline{$^{29}$University of California, Irvine, California 92697}
\centerline{$^{30}$University of California, Riverside, California
                   92521}
\centerline{$^{31}$Florida State University, Tallahassee, Florida
                   32306}
\centerline{$^{32}$University of Hawaii, Honolulu, Hawaii 96822}
\centerline{$^{33}$Fermi National Accelerator Laboratory, Batavia,
                   Illinois 60510}
\centerline{$^{34}$University of Illinois at Chicago, Chicago,
                   Illinois 60607}
\centerline{$^{35}$Northern Illinois University, DeKalb, Illinois
                   60115}
\centerline{$^{36}$Northwestern University, Evanston, Illinois 60208}
\centerline{$^{37}$Indiana University, Bloomington, Indiana 47405}
\centerline{$^{38}$University of Notre Dame, Notre Dame, Indiana
                   46556}
\centerline{$^{39}$Iowa State University, Ames, Iowa 50011}
\centerline{$^{40}$University of Kansas, Lawrence, Kansas 66045}
\centerline{$^{41}$Kansas State University, Manhattan, Kansas 66506}
\centerline{$^{42}$Louisiana Tech University, Ruston, Louisiana 71272}
\centerline{$^{43}$University of Maryland, College Park, Maryland
                   20742}
\centerline{$^{44}$Boston University, Boston, Massachusetts 02215}
\centerline{$^{45}$Northeastern University, Boston, Massachusetts
                   02115}
\centerline{$^{46}$University of Michigan, Ann Arbor, Michigan 48109}
\centerline{$^{47}$Michigan State University, East Lansing, Michigan
                   48824}
\centerline{$^{48}$University of Nebraska, Lincoln, Nebraska 68588}
\centerline{$^{49}$Columbia University, New York, New York 10027}
\centerline{$^{50}$New York University, New York, New York 10003}
\centerline{$^{51}$University of Rochester, Rochester, New York 14627}
\centerline{$^{52}$State University of New York, Stony Brook,
                   New York 11794}
\centerline{$^{53}$Brookhaven National Laboratory, Upton, New York
                   11973}
\centerline{$^{54}$Langston University, Langston, Oklahoma 73050}
\centerline{$^{55}$University of Oklahoma, Norman, Oklahoma 73019}
\centerline{$^{56}$Brown University, Providence, Rhode Island 02912}
\centerline{$^{57}$University of Texas, Arlington, Texas 76019}
\centerline{$^{58}$Texas A\&M University, College Station, Texas
                   77843}
\centerline{$^{59}$Rice University, Houston, Texas 77005}
\centerline{$^{60}$University of Washington, Seattle, Washington
                   98195}

\clearpage

\normalsize
\setlength{\parskip}{0.8ex plus0.5ex minus0.2ex}

%---------------------------------------------------------------------
%\section*{Introduction}

The top quark is the charge $+2/3$ weak-isospin partner of the bottom
quark in the third generation of fermions of the standard model
(SM). It is extremely massive at
{\mbox{$174.3\pm5.1$~GeV~\cite{topmass}}, and, with an expected width
of $1.5$~GeV~\cite{topwidth}, it decays before hadronization almost
exclusively into a $W$~boson and a $b$~quark. At the Tevatron {\ppbar}
collider, most top quarks are pair-produced via the strong interaction
through an intermediate gluon. This was the mode used in its
observation~\cite{topdiscovery} and subsequent studies of its
properties, including measurements of the {\ttbar} production cross
section of {\mbox{$5.9\pm1.7$~pb} by the {\dzero}
collaboration~\cite{d0-tt-xsec}, and {\mbox{$6.5^{+1.7}_{-1.4}$~pb} by
the CDF collaboration~\cite{cdf-tt-xsec}. A second production mode is
predicted to exist, where top quarks are created singly through an
electroweak $Wtb$ vertex~\cite{sintoptheory}. Many processes beyond
the SM can boost the single top quark cross section~\cite{yuan}. In
the absense of a cross section excess, measurement of the electroweak
production of single top quarks could provide the magnitude of the CKM
matrix element {\Vtb}~\cite{measVtb}, since the cross section is
proportional to {\Vtbsq}. In this Letter, we describe a search for
single top quarks at the Tevatron, using data collected from
1992--1995 at a {\ppbar} center-of-mass energy of 1.8~TeV.

%\section{Single Top Quark Production}

\begin{figure}[!h!tbp]
\vspace{0.2 in}
\centerline
{\protect\psfig{figure=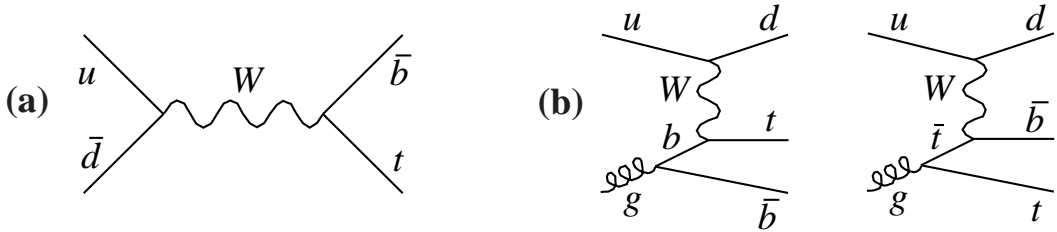,width=5.5in}}
\begin{center}
\begin{minipage}{5.5 in}
\caption[fig1]{Leading order Feynman diagrams for single top quark
production at the Tevatron, where (a) shows the $s$-channel mode, and
(b) the $t$-channel mode.}
\label{fig1}
\end{minipage}
\end{center}
\vspace{-0.1 in}
\end{figure}

%%\subsection{Three Production Modes}

The standard model predicts three modes for the production of single
top quarks at a hadron collider. The first is the $s$-channel process
$q^{\prime}\bar{q}{\rar}tb$, illustrated in Fig.~\ref{fig1}(a). For a
top quark mass $m_t$ of 175~GeV, this has a cross section calculated
at next-to-leading-order (NLO) of
{\mbox{$0.73\pm0.10$~pb}~\cite{willenbrock-s}}. Following the decay of
the top quark, these events contain a $W$~boson and two $b$~quarks
that hadronize into two central jets with high transverse momentum
($p_T$). The second production mode, shown in Fig.~\ref{fig1}(b) and
sometimes referred to as ``$W$-gluon fusion'', is a $t$-channel
process, $q^{\prime}g{\rar}tqb$. The NLO cross section is
{\mbox{$1.70\pm0.24$~pb~\cite{willenbrock-t}}. This process produces a
$W$~boson, a forward light-quark jet, and two central $b$~jets, one
with high~$p_T$ and the other with low $p_T$. We have searched for
both production modes, with decay of the $W$~boson into $e\nu$ or
$\mu\nu$, and identification of a $b$~jet via a tagging muon. A third
mode occurs via both the $s$-channel and $t$-channel, $bg{\rar}tW$,
with a final state containing two $W$~bosons and a single $b$~jet. The
leading-order cross section for this process is only
0.15~pb~\cite{heinson}, and, with $\approx90$~pb$^{-1}$ of available
data, there is no possibility of separating it from
background. Throughout this Letter, we use ``$tb$'' to refer to both
$t\bar{b}$ and the charge-conjugate process $\bar{t}b$, and ``$tqb$''
to both $tq\bar{b}$ and~$\bar{t}\bar{q}b$.

%\section{Detector and Data}

The {\dzero} detector~\cite{d0nim} has three major components: a
central tracking system including a transition radiation detector
(TRD), a uranium/liquid-argon calorimeter, and a muon spectrometer.
For the measurement in the electron channel, we use
$91.9\pm4.1$~pb$^{-1}$ of data collected with a trigger that required
an electromagnetic (EM) energy cluster in the calorimeter, a jet, and
missing transverse momentum ({\met}). For events passing the final
selection, the efficiency of the trigger is 90--$93\%$, depending
on the location of the EM cluster in the calorimeter. In the muon
channel, we use $88.0\pm3.9$~pb$^{-1}$ of data acquired with several
triggers, which required {\met} or a muon with a jet. The combined
efficiency of these triggers is 96--$99\%$. A third data sample,
obtained with a trigger requiring just three jets, is used for
measuring one of the backgrounds. Since the multijet cross section is
very large, this trigger was prescaled, and we have 0.8~pb$^{-1}$ of
such data. Each of the three samples contains approximately one
million events.

%\section{Event Reconstruction}

%%\subsection{Electron Identification}

To determine whether an EM energy cluster was generated by an
electron, we require it to be isolated from other activity in the
calorimeter and use a five-variable likelihood function to
discriminate electrons from background. This likelihood includes the
fraction of cluster energy contained in the EM region of the
calorimeter ($>90\%$ for electrons), the cluster shape (it must
resemble an electron and not a pion), the presence of a well-matched
track between the cluster and a primary {\ppbar} interaction vertex
(to discriminate against photons), the $dE/dx$ energy loss along the
track (consistent with a single particle and not from a photon
conversion into a pair of charged particles), and the TRD response
(matching that of an electron and not a pion). An electron is then
required to have transverse energy $E_T>20$~GeV, and to be within the
optimal region of the calorimeters with detector pseudorapidity
${\mdeta}<1.1$ or $1.5<{\mdeta}<2.5$~\cite{deta-definition}. When an
electron is isolated, it is more likely to have originated from the
decay of a $W$~boson than from a $b$~hadron. The efficiency of the
combined electron identification requirements is $\approx60\%$.

%%\subsection{Jet Identification}

Jets, reconstructed with a cone algorithm of radius
$R=0.5$~\cite{jet-definition}, must fail the electron requirements.
The jet with highest transverse energy is required to have
$E_T>15$~GeV and ${\mdeta}<3.0$. The second jet has to have
$E_T>10$~GeV and ${\mdeta}<4.0$. Other jets in the event are counted
if they have $E_T>5$~GeV and ${\mdeta}<4.0$. We set the $E_T$
thresholds low and the ${\mdeta}$ region wide to maximize acceptance
for signal; however, the efficiency to reconstruct jets close to the
5~GeV threshold is low.

%%\subsection{Muon Identification}

We identify a muon by the pattern of hits in the spectrometer drift
tubes, and require an impact parameter $<20$~cm between the
spectrometer track and the primary vertex, a matching track in the
calorimeter consistent with a minimum-ionizing particle, a matching
central track, a signal in the scintillators surrounding the
spectrometer within $\pm12$~ns of the beam-crossing time, and
penetration through one of the spectrometer toroids for momentum
analysis. Most of these requirements are designed to reject cosmic
rays and particles backscattered from the beamline magnets. Muons must
be within the central region of the spectrometer, with
${\mdeta}<1.7$. A muon is called ``isolated'' if $\Delta
R\left(\mu,{\rm jet}\right) \ge 0.5$~\cite{dR-definition} for all jets
with $E_T>5$~GeV. An isolated muon must have $p_T>20$~GeV and is
attributed to the decay of a $W$~boson. A ``tagging'' muon has $\Delta
R<0.5$ and $p_T>4$~GeV. It is attributed to the semileptonic decay of
a $b$~hadron in a jet, and thus identifies a $b$~jet. The efficiency
of the combined muon identification requirements is $\approx44\%$ for
isolated muons.

%%\subsection{Neutrino Identification}

Because a leptonically-decaying $W$~boson is supposed to be present in
each signal event, we require ${\met}>15$~GeV as evidence of a
neutrino.

\clearpage

%\section{Analysis Overview}

We use the NLO single top quark production cross sections to estimate
that about 66~\mbox{$s$-channel} and 153 $t$-channel events were
produced at {\dzero} during Run~1. Of these, we expect that about
15~$s$-channel and 35 $t$-channel events passed the trigger
requirements and were recorded for analysis.

%\section{Signal Selection}

%%\subsection{Baseline Set of Cuts}

Our analysis starts with a simple baseline selection of events that
pass the triggers and have at least one reconstructed electron or
isolated muon, and at least two jets with $E_T>5$~GeV and
${\mdeta}<4.0$. For the results presented in this Letter, we also
demand at least one tagging muon (``$/\mu$''), to indicate the
possible presence of a $b$~jet. These minimal requirements reduce the
$\approx1$~million events in each channel to 116~{\ejm} events and
110~{\mjm} events. The acceptance for single top quark events for
these selections is \mbox{0.2--$0.3\%$} per channel, which should
yield $\approx1$ tagged event ($tb$ and $tqb$, with electron and muon
$W$ decays combined). The expected number of events is small because
the probability to identify at least one tagging muon in a single top
quark event is only 6--$11\%$. After these selections, $90\%$ of the
background in the electron channel is from QCD multijet production
with a jet misidentified as an electron, $5\%$ from {\ttbar} events,
and $5\%$ from $W$+jets (including $WW$ and $WZ$ diboson events),
where about two thirds of the $W$+jets events have a light quark or
gluon jet with a false tagging muon, and a quarter of the tagging
muons are from $c$~quark decays. In the muon channel, the background
is $8\%$ from $W$+jets events, $6\%$ from QCD {\bbbar} events where a
muon from a $b$~decay mimics an isolated muon, and $4\%$ from {\ttbar}
events. The remaining $82\%$ of the background is from QCD multijet
events with a coincident cosmic ray or beam-halo particle
misidentified as an isolated muon.

%%\subsection{Loose Set of Cuts}

Next, we apply a set of loose criteria to remove mismeasured events
and to reject backgrounds that do not have the same final-state
characteristics as our signal. We reject events with more than one
isolated lepton and any isolated photons. We remove events with {\met}
close to 15~GeV and aligned with or opposite to a jet, or opposite an
electron or isolated muon. We also reject events that have muons with
clearly mismeasured $p_T$. We require two, three, or four jets. To
remove the remaining contamination from cosmic rays in the isolated
muon channel, we reject events where the isolated muon and tagging
muon are back-to-back; in particular, we require $\Delta\phi({\rm
isol}\mu,{\rm tag}\mu)<2.4$~rad. These criteria, together with the jet
$E_T$ and {\mdeta} requirements and the {\met} threshold, reject
$86\%$ of the baseline multijet ``electron'' events, $95\%$ of the
cosmic ray and misreconstructed isolated ``muon'' events, $90\%$ of
the {\bbbar}~``isolated'' muon events, $27\%$ of the $W$+jets events
in the electron channel, $81\%$ in the muon channel, and 55--$73\%$ of
the {\ttbar} background. The signal acceptances are reduced by
14--$51\%$. There remain 21~{\ejm} and 8~{\mjm} candidates in the
data.

%%\subsection{Tight Sets of Cuts}

Based on independent studies (see below), we apply the following
requirements to obtain the best significance of signal over
square-root of background in each channel:

\vspace{0.02 in}
\noindent Electron Channel
\vspace{-0.09 in}
\begin{itemize}
\item $E_T({\rm jet1}) + E_T({\rm jet2}) + E_T(e) + {\met}
\; > \; 125$~GeV
\item $E_T({\rm jet3}) + 5 \times E_T({\rm jet4}) \; < \; 47$~GeV
\item $E_T({\rm jet1}) + 4 \times {\met} \; > \; 155$~GeV
\end{itemize}

\clearpage

\noindent Muon Channel
\vspace{-0.09 in}
\begin{itemize}
\item $E_T({\rm jet1}) + E_T({\rm jet2}) + E_T({\rm jet3}) +
E_T({\rm jet4}) \; > \; 70$~GeV
\item $E_T({\rm jet3}) + 5 \times E_T({\rm jet4}) \; < \; 47$~GeV
\end{itemize}

The first criterion in each set was chosen by studying reconstructed
{\comphep}~\cite{comphep} Monte Carlo (MC) $W$+jets events, the second
by examining {\herwig}~\cite{herwig} {\ttbar}~MC events, and the third
variable in the electron channel was determined from studies of QCD
multijet data. The distributions were compared with signal MC from
{\comphep}. The cutoffs were optimized on combined samples of untagged
and tagged events. Figure~{\ref{fig2} shows the distribution of the
second variable, designed to minimize {\ttbar} background, for
electron and muon events combined after all other selections have been
applied.

\begin{figure}[!h!tbp]
\centerline
{\protect\psfig{figure=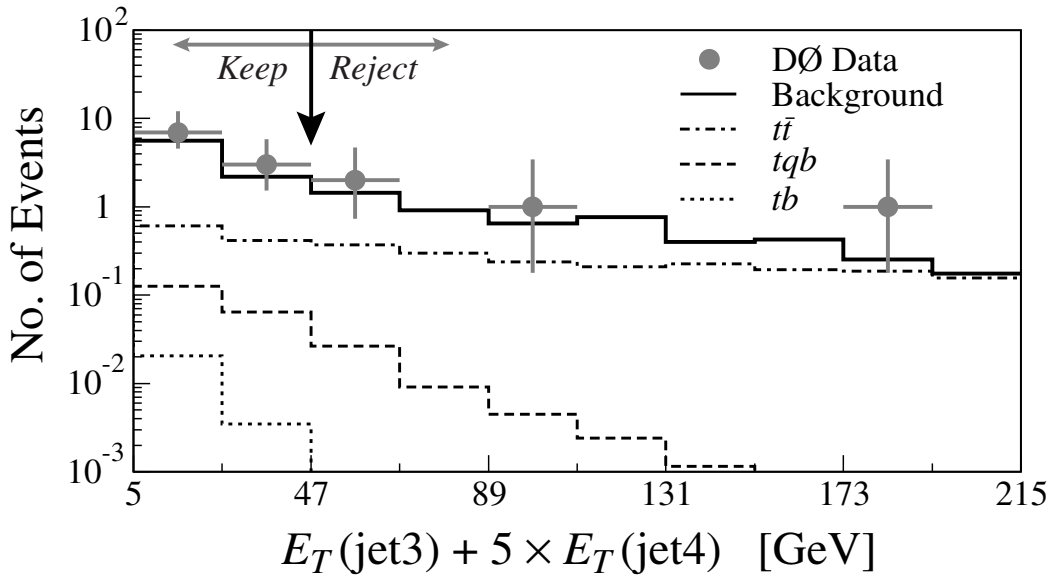,width=5.5in}}
\begin{center}
\caption[fig2]{Variable used to reject {\ttbar} background.}
\label{fig2}
\vspace{-0.2 in}
\end{center}
\end{figure}

After final selections, there is no evidence of an excess of signal
over background, and we therefore use the results to set limits on the
$s$-channel and $t$-channel single top quark cross sections. To do
this, we must first determine the signal acceptance and the background
in each channel.

%\section{Acceptance and Background Measurements}

We obtain the signal acceptances using MC samples of $s$-channel and
$t$-channel single top quark events from the {\comphep} event
generator, with the {\pythia} package~\cite{pythia} used to simulate
fragmentation, initial-state and final-state radiation, the underlying
event, and leptonic decays of the $W$~boson. The MC events are
processed through a detector simulation program based on the
{\geant}~\cite{geant} package and a trigger simulation, and are then
reconstructed. We apply all selections directly to the reconstructed
MC events, except for several particle identification criteria, which
we correct using factors measured in other {\dzero}
data. Table~\ref{table1} shows the acceptance for single top quark
events after all selection requirements and corrections.

\begin{table}[!h!btp]
\begin{center}
\begin{minipage}{4 in}
\caption[tab1]{Signal acceptances (as percentages of the total cross
sections), and numbers of events expected to remain after application
of all selection criteria.\vspace{0.1 in}}
\label{table1}
\begin{tabular}{lcc}
            & Electron Channel  & Muon Channel
\vspace{0.02 in}                                        \\
\hline
\vspace{0.05 in}
 & \multicolumn{2}{c}{\underline{Acceptances}}
\vspace{0.02 in}                                        \\
~~$tb$      & $(0.255\pm0.022)\%$ & $(0.112\pm0.011)\%$ \\
~~$tqb$     & $(0.168\pm0.015)\%$ & $(0.083\pm0.008)\%$
\vspace{0.05 in}                                        \\
 & \multicolumn{2}{c}{\underline{Numbers of Events}}
\vspace{0.02 in}                                        \\
~~$tb$      &   $0.18\pm0.03$   &   $0.08\pm0.01$       \\
~~$tqb$     &   $0.28\pm0.05$   &   $0.13\pm0.03$       \\
~~$W$+jets  &   $5.59\pm0.64$   &   $1.12\pm0.17$       \\
~~QCD       &   $5.92\pm0.58$   &   $0.40\pm0.09$       \\
~~{\ttbar}    &   $1.14\pm0.35$   &   $0.45\pm0.14$
\vspace{0.02 in}                                        \\
Total Bkgd  &  $12.65\pm0.93$   &   $1.97\pm0.24$
\vspace{0.02 in}                                        \\
Data        &       $12$        &       $5$
\end{tabular}
\end{minipage}
\end{center}
\end{table}

The acceptance for {\ttbar} pairs is calculated in a manner similar to
that for signal and then converted to a number of events using the
integrated luminosity for each channel and {\dzero}'s value of the
{\ttbar} cross section~\cite{d0-tt-xsec}.

The QCD multijet background with a jet misidentified as an electron is
measured using multijet data. The events are weighted by the
probability that a jet mimics an electron for each jet that passes the
electron $E_T$ and {\mdeta} requirements. These probabilities are
determined from the same multijet sample, but for ${\met}<15$~GeV, and
are found to be $(0.0160\pm 0.0016)\%$ for ${\mdeta}<1.1$, and
$(0.0622\pm 0.0048)\%$ for ${\mdeta}>1.5$. We normalize the integrated
luminosity of the multijet sample so as to match the data sample used
in the search for the signal, and correct for a small difference in
trigger efficiency between the two samples.

The QCD {\bbbar} background arises when both $b$~quarks decay
semileptonically to a muon, and one muon is misidentified as
isolated. There are two ways for such events to mimic the
signal. First, one of the $b$~jets may not be reconstructed, and its
muon can therefore appear to be isolated. Second, a muon can be
emitted wide of its jet and be reconstructed as an isolated muon. The
background from each source is measured using data collected with the
same triggers as used for the muon signal. The events are required to
pass all selections, except that the muon, which otherwise passes the
isolated muon requirements, is within a jet. Events with truly
isolated muons are excluded. Each event is then weighted by the
probability that a nonisolated muon is reconstructed as an isolated
one. This probability is measured using the same data sample, but for
${\met}<15$~GeV, and found to be $(2.94\pm0.53)\%$ for the case of a
``lost~jet,'' and $(1.38\pm0.25)\%$ for a ``wide~$\mu$,'' for muons
with $p_T < 32$~GeV. The probabilities are parametrized as a function
of the muon $p_T$. We calculate a weighted average of the two results
to obtain the number of expected background events.

\clearpage

The background from $W$+jets is estimated by applying a set of
tag-rate functions to untagged signal candidates that pass all final
event selections. These tag-rate functions are measured using multijet
data and correspond to the relative probability that a jet of given
$E_T$ and {\deta} has a tagging muon, for two run periods when the
muon chambers had different operating efficiencies. We then correct
the samples for a small difference in trigger efficiency between
untagged and tagged events. We also correct the muon channel by a
factor of \mbox{$0.688\pm0.034$} to account for the effect of the
$\Delta\phi$ cutoff used to minimize cosmic ray backgrounds, a
selection that cannot be applied directly. Finally, to avoid double
counting, we subtract the fraction of events expected from {\ttbar}
and QCD backgrounds and single top quark signals. The remaining
fraction of $W$+jets in the untagged signal candidates is
\mbox{66--$92\%$}, depending on the location of the electron or
isolated muon.

The numbers of events expected for the two signals and three
backgrounds are shown in Table~\ref{table1}, together with the final
numbers of events in the candidate data samples, for the electron and
muon channels.

%\section{Cross Section Limits}

To calculate limits on the cross sections for single top quark
production in the $s$-channel and $t$-channel modes, we use the
numbers of observed events, the signal acceptances and backgrounds,
and the integrated luminosities. Covariance matrices are used to
describe the correlated uncertainties on these quantities. We use a
Bayesian approach, with a flat prior for the single top quark cross
section and a multivariate Gaussian prior for the other quantities. We
calculate the likelihood functions in each decay channel and combine
them to obtain the following $95\%$ confidence level upper limits:

\vspace{-0.1 in}
\begin{itemize}
\item $\sigma({\ppbar}{\rargap}tb+X) < 39$~pb
\item $\sigma({\ppbar}{\rargap}tqb+X) < 58$~pb
\end{itemize}
\vspace{-0.1 in}

%\section{Summary}

To conclude, we have searched for electroweak production of single top
quarks and find no evidence for such production. We set upper limits
on the cross sections for $s$-channel production of $tb$ and
$t$-channel production of $tqb$. The limits are consistent with
expectations from the standard model.

%\section*{Acknowledgments}

We are grateful to T.~Stelzer, Z.~Sullivan, S.~Willenbrock, and
C.-P.~Yuan for valuable discussions on theoretical aspects of the
analysis. We thank the staffs at Fermilab and at collaborating
institutions for contributions to this work, and acknowledge support
from the Department of Energy and National Science Foundation (USA),
Commissariat \` a L'Energie Atomique and CNRS/Institut National de
Physique Nucl\'eaire et de Physique des Particules (France), Ministry
for Science and Technology and Ministry for Atomic Energy (Russia),
CAPES and CNPq (Brazil), Departments of Atomic Energy and Science and
Education (India), Colciencias (Colombia), CONACyT (Mexico), Ministry
of Education and KOSEF (Korea), CONICET and UBACyT (Argentina),
A.P.~Sloan Foundation, and the A.~von~Humboldt Foundation.


\begin{thebibliography}{99.}
\bibitem{deta-definition}
The detector pseudorapidity {\deta} is defined as the pseudorapidity
of the calorimeter cluster (electron or jet) or central track (muon)
with respect to the center of the {\dzero} detector, where
${\deta}=-\ln\tan({\dtheta}/2)$, and {\dtheta} is the polar angle
relative to the proton direction. The pseudorapidity $\eta$ and polar
angle $\theta$ are defined with respect to the primary interaction
point.
\bibitem{jet-definition}
The cone radius of a jet is defined as
$R=\sqrt{(\Delta\phi)^2+(\Delta\eta)^2}$, where $\Delta\phi$,
$\Delta\eta$ specify the width of the jet, and $\phi$ is the azimuthal
angle around the beam axis.
\bibitem{dR-definition}
The opening angle between a muon and jet is defined as \hfill \break
$\Delta R=\sqrt{\left[\Delta\phi\left({\mu},{\rm jet}\right)\right]^2
         +\left[\Delta\eta\left({\mu},{\rm jet}\right)\right]^2}$.
\end{thebibliography}
\end{document}